\documentclass[a4paper,11pt]{article}
\usepackage{jinstpub} 

\usepackage{subcaption} 

\title{\boldmath Ring oscillator performance of the ATLAS inner tracker pixel readout chip}

\author[a,1,2]{Yahya Khwaira \note{Corresponding author.}\note{Now at Laboratoire de Physique Nucléaire et des Hautes Énergies (LPNHE), CNRS - Sorbonne Université, Université Paris Cité, Paris, France}}
\author[a]{Abdenour Lounis}
\author[a]{Maurice Cohen-Solal}
\author[b]{Mohsine Menouni}
\author[b]{Pierre Barrillon}
\author[b]{Denis Fougeron}

\affiliation[a]{Laboratoire de Physique des 2 Infinis Irène Joliot Curie (IJC Lab), Centre National de la Recherche Scientifique (CNRS), Paris Saclay University, Orsay, France}
\affiliation[b]{Centre de Physique des Particles de Marseille (CPPM), Centre National de la Recherche Scientifique (CNRS), Aix-Marseille University, 13288, Marseille, France}

\emailAdd{ykhwaira@cern.ch}

\abstract{This paper presents experimental and simulation data to characterize the Ring Oscillators (RO) produced in 65-nm CMOS technology for the next promising generation of readout chips for the pixel detector in the Inner Tracker (ITk) at the ATLAS experiment at CERN. To enable a better understanding of the RO block embedded in ITkPixV1.1 single chip card (SCC), tests at various temperatures, voltages, accumulated total ionizing dose (TID) with X-ray irradiation, and high-temperature annealing will be presented. The objective of this study is to examine the RO output dependency based on different variable conditions and provide simulation data using Cadence, an electronic design automation (EDA) software to validate the experimental outcomes.}


\begin{document}
\maketitle
\flushbottom

\section{Introduction}
\label{sec:intro_of_intro}
A ring oscillator (RO) is a device that can be integrated into readout chips for high-energy physics (HEP) experiments for radiation monitoring. The oscillation frequency in the RO is an element in understanding the behavior of external influences on the chip due to any change in temperature, voltage, or irradiation during the experimental data collection of the ATLAS experiment \cite{TheATLASCollaboration2008}. Hence, it provides a clue about the external changes on the overall detector module and monitoring the chip configuration. 

The motivation of this study is to test ITkPixV1.1, a readout chip prototype developed within the RD53 collaboration \cite{LODDO2024169682} for the pixel detector at the ATLAS experiment at CERN. Due to the foreseen High Luminosity Large Hadron Collider (HL-LHC) upgrade, an imposed higher luminosity of $5 \times 10^{-34}\,\mathrm{cm^{-2}\,s^{-1}}$ and an expected 1~Grad TID accumulated over 10~years \cite{meng2021atlas} at the innermost detector layer in ATLAS, presents significant challenges on the electronics that will suffer from long-term radiation effects \cite{yah_measurment} and degradation over time, such as threshold shifts, increased leakage current, propagation delay, and power consumption \cite{MOLL199987}.

The RO block device, which is the target of this study, will degrade over time as the propagation delay $T_{pd}$ of the signal in the RO gates will increase due to the radiation effects \cite{Schlenvogt2013SimulationOT}. As the RO is being used as a radiation sensor, it indicates the overall performance of the readout chip digital cells. A harsh operation environment at the ITk at $-40^{\circ}\mathrm{C}$ and a harsh radiation environment impose the need to test and characterize the RO device.

 In this paper, we present the methodology for testing the RO block and acquiring valuable information under controlled conditions to study the variables affecting RO frequency: temperature, voltage, total ionizing dose (TID), and high-temperature annealing after irradiation. The ITkPixV1.1 SCC was tested in a climate chamber across various temperatures while controlling the bias voltage, then irradiated with an X-ray tube delivering 16~krad/min over 25~days to reach a total dose of 520~Mrad, followed by a 62-day annealing period to observe performance recovery at high temperature. Additionally, Cadence simulations were performed to support the experimental study and characterize the RO’s frequency dependencies.

The paper is organized as follows: In section~\ref{sec:ROSC_intro}, we describe the RO device and highlight the main dependencies in our scope. section~\ref{sec:ROSC_in_SCC} explains the RO block in the testing chip. section~\ref{sec:Cadence} handles circuit simulation with Cadence. In section~\ref{sec:Experimental_Validation}, we describe the X-ray irradiation setup, testing methodology, and analysis, with pre-irradiation data validated by simulation, and we highlight the RO output evolution during annealing. Finally, section~\ref{sec:Conc} summarizes the study's conclusions.

\section{The Ring oscillator device}
\label{sec:ROSC_intro}
A RO is a type of oscillator circuit that produces a periodic waveform output. It consists of an odd number of inverting gates, such as NAND or NOR gates, arranged in a closed loop. The output of the last gate in the chain is fed back to the input of the first gate, creating a feedback loop.

\begin{figure}
    \centering

    \includegraphics[width=0.8\linewidth]{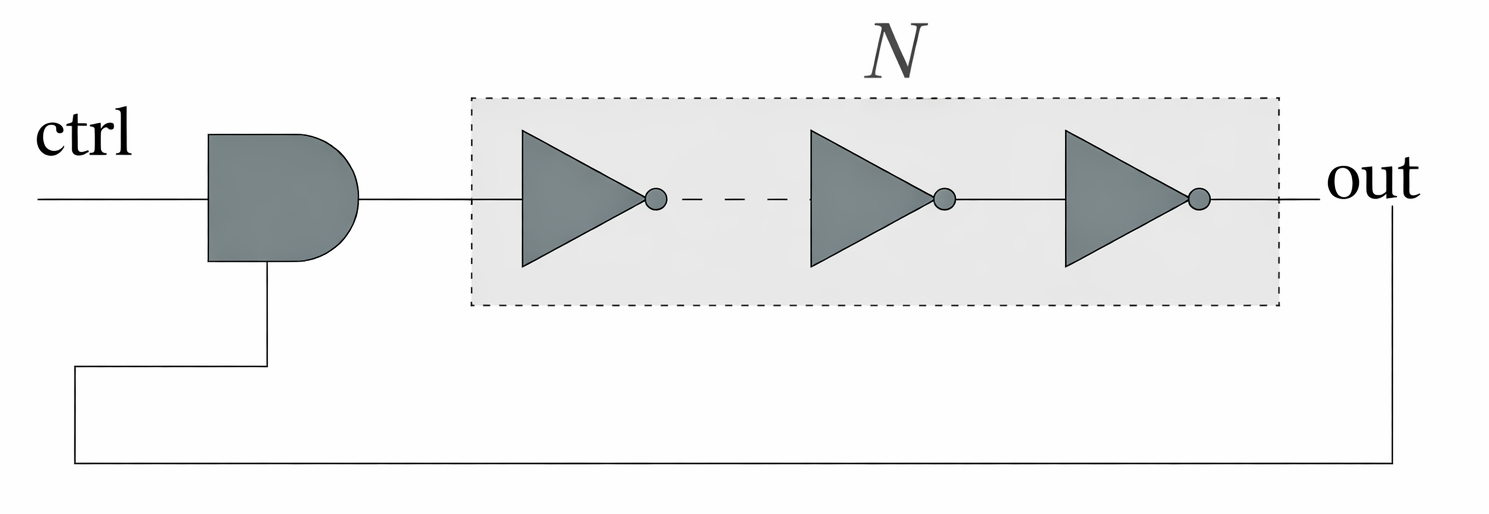}
    \caption{Basic RO circuit with N number of gates.}
    \label{fig:ROSC_Diagram}
\end{figure}
The time it takes for the signal to complete one cycle of the ring is called the oscillator period. The frequency of the oscillator, which is the number of cycles per unit time, is the inverse of the period and can be written as follows:
\begin{equation}
f_{RO} = \frac{1}{2NT_{pd}},
\label{eq:freq}
\end{equation}
where $f_{RO}$ is the oscillation frequency, $N$ is the number of equivalent gates in the ring, $T_{pd}$ is the propagation delay through each gate, and the factor 2 represents the fact that a full cycle requires both a low-to-high and a high-to-low transition.
Moreover, the propagation delay through each gate is dependent on the characteristics of the gate, such as its size and threshold voltage. The gate propagation delay $T_{pd}$ can be approximated using the following equation \cite{10.5555/1841628}:
\begin{equation}
T_{pd} \approx \frac{\left(\frac{L}{W}\right)C_{L}}{\mu C_{OX}V_{DSat}}
= \frac{\left(\frac{L}{W}\right)C_{L}}{\mu C_{OX} P_{v}\left(V_{dd}-V_{th}\right)^{\sigma/2}},
\label{eq:big_eq}
\end{equation}
where $P_v$ and $\sigma$ are technology-dependent constants empirically derived from the I-V characteristics, $C_{L}$ is the load capacitance seen by the gate output, $W$ and $L$ are the width and length of the transistor gate, $V_{dd}$  is the supply voltage, and $V_{th}$ is the threshold voltage of the transistors and $C_{OX}$ is the oxide gate capacitance. Note that this equation is a simplified approximation and does not consider all of the parasitic capacitances and resistances that interconnecting the gates and may affect the gate delay. It can be noticed as well that increasing $V_{dd}$ results in lower propagation delays and higher frequency outputs.

\subsection{Temperature effects}
Temperature changes can affect the frequency of the RO because of changes in the material properties used in fabrication. The speed of charge carriers $\upsilon$  inside the transistor channels is proportional to the electric field $E$ applied between the source and drain. Furthermore, transistor carrier mobility as a function of temperature change can be approximated as \cite{1487607}:  
\begin{equation}
\mu(T) = \mu(T_0)\left(\frac{T}{T_0}\right)^{k_m},
\qquad k_m \approx -1.2 \text{ to } -2,
\label{eqn:mobilityrosc}
\end{equation}
where $T$ and $T_0$ are the absolute and reference temperatures respectively, and $k_m$ is a fitting parameter. Hence, as the temperature increases, the mobility of charge carriers $\mu$ will decrease, increasing the propagation delay $T_{pd}$ as noted in equation~\ref{eq:big_eq}, due to slower transistor switching speeds. Thus, this increase in propagation delay  $T_{pd}$ leads to a decrease in the RO frequency.

\subsection{TID effects}
Radiation-induced damage by high-energy particles such as protons, neutrons, electrons, or heavy nuclei, as expected due to proton–proton collisions in the HL-LHC, imposes a significant challenge for the operation of the pixel detector modules \cite{CERN-LHCC-2017-021}. Over the long run, the TID damage of 1~Grad accumulated over 10~years of operation, can cause severe changes in the threshold voltage and carrier mobility leading to oxide charging that affects the  performance of the Pixel readout chip.  In our paper, we study the X-ray photon interaction with silicon transistor bulk using X-ray tube and cover the related major TID damage mechanisms. Trapped charges in the gate oxide and at the silicon-oxide interface lead to a shift in the threshold voltage $\Delta V_{th}$, primarily driven by the buildup of oxide-trapped charge $\Delta V_{ot}$ and interface-trapped charge $\Delta V_{it}$ \cite{osti_5646360, ZHANG2021107951}. The combined shift alters the transistor switching behavior and, consequently, the propagation delay in logic gates, resulting in a measurable shift in the RO oscillation frequency. Monitoring this frequency shift enables an indirect estimation of the absorbed dose during irradiation \cite{10.1007/1-4020-4367-8_24, 4033191}. Therefore, in the context of measuring the oscillation frequency, the deviation of this magnitude from the nominal value can be interpreted as the absorbed dose:
\begin{equation}
\mathrm{TID}\ \rightarrow\ \Delta V_{ot} + \Delta V_{it} = \Delta V_{th}\ \rightarrow\ f_{RO}.
\end{equation}
After irradiation, thermal annealing can partially recover device performance by enabling the release (detrapping) of radiation-induced charges from oxide and interface traps. This thermally activated process reduces the accumulated threshold voltage shift $\Delta V_{th}$, thereby restoring the RO frequency closer to its pre-irradiation value. The detrapping behavior follows first-order kinetics and depends on temperature and trap characteristics \cite{9857702}.

\begin{figure}
    \centering
    \includegraphics[width=0.75\linewidth]{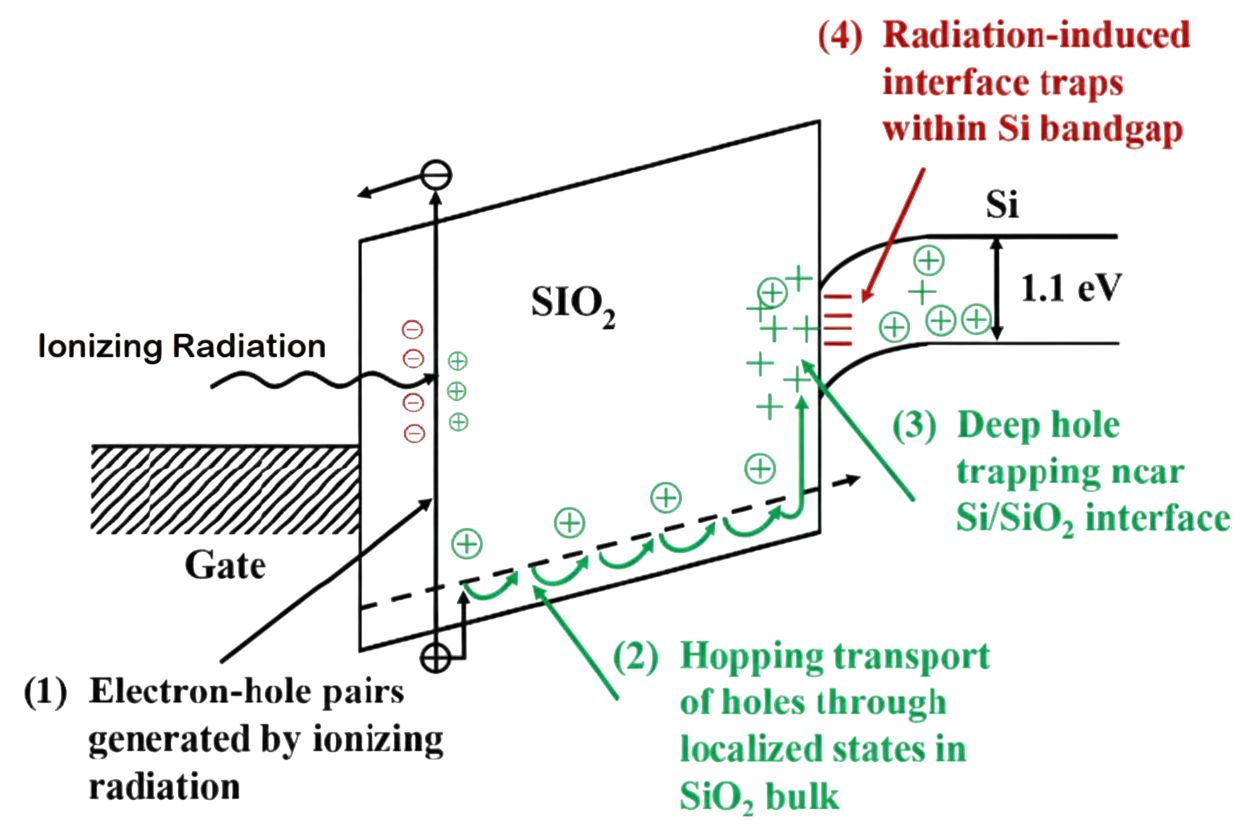}
    \caption{Fundamental radiation-induced electron-hole pair generation with bulk hole and interface trapping.}
    \label{fig:enter-label}
\end{figure}

\section{Ring Oscillator in ITkPixV1.1}
\label{sec:ROSC_in_SCC}
The RO block implemented in the ITkPixV1.1 SCC includes multiple gate drive strength variants \cite{Garcia-Sciveres:2665301}. In this work, two strength configurations are studied: Str-0 and Str-4. These variants share the same logical topology and channel length, but differ in their effective transistor channel width, which is implemented by selecting a different number of transistor fingers in the layout. The Str-4 configuration uses a larger number of fingers, resulting in a wider effective channel width and higher drive strength compared to Str-0.

This difference directly impacts the propagation delay and radiation tolerance of the RO gates, as indicated in equation~\ref{eq:big_eq}. In addition to drive strength variations, the ITkPixV1.1 SCC implements ROs constructed from different digital logic gate types, including inverters, NAND, NOR, AND, and OR gates. Each RO is formed by cascading an odd number $N$ of identical gates of a given type, resulting in distinct RO lengths and logic topologies. This diversity in gate type and RO length is intentionally designed to enable a systematic study of radiation-induced effects across different logic functions and propagation paths.
By comparing ROs with different gate types, drive strengths, and numbers of stages, it is possible to disentangle the impact of irradiation on transistor topology, effective channel width, and cumulative propagation delay \cite{Jiang2011}. This approach provides improved insight into how total ionizing dose and annealing affect different digital building blocks used in the readout chip.

\section{Cadence Simulation}
\label{sec:Cadence}
To evaluate the performance of the RO circuit, simulations were performed to extract key parameters such as frequency and propagation delay under varying conditions. The simulation workflow includes schematic design, layout validation, and electrical verification across different operating scenarios.

Two key parameters were varied independently: supply voltage and temperature. The supply voltage $V_{dd}$  was swept from 0.9~V to 1.2~V in 0.1~V increments, matching the typical operating range of ITkPixV1.1. Temperature was varied from $-20^{\circ}\mathrm{C}$ to $60^{\circ}\mathrm{C}$, consistent with the expected experimental range. As this range already captures the dominant temperature-dependent behavior of the RO, experimental data extend to $-40^{\circ}\mathrm{C}$ and show a consistent trend. During each simulation sweep, other factors were held constant to isolate the effect under study. While real-world behavior includes combined effects such as TID, these were decoupled in simulation to allow individual analysis of their impact on propagation delay. 

\section{Experimental Validation}
\label{sec:Experimental_Validation}
To validate the simulation results obtained with Cadence, an irradiation campaign has been conducted at the Institute Materials Microelectronics Nanoscience of Provence (IM2NP). In this section, we introduce the relevant pre-irradiation tests using the SCC hosted inside a climate chamber and then explain the irradiation campaign plan. Moreover, post-irradiation annealing at high temperatures will be presented to deduce the RO performance closely to what is expected in the HL-LHC. 

\begin{figure}
    \centering
    \includegraphics[width=0.8\linewidth]{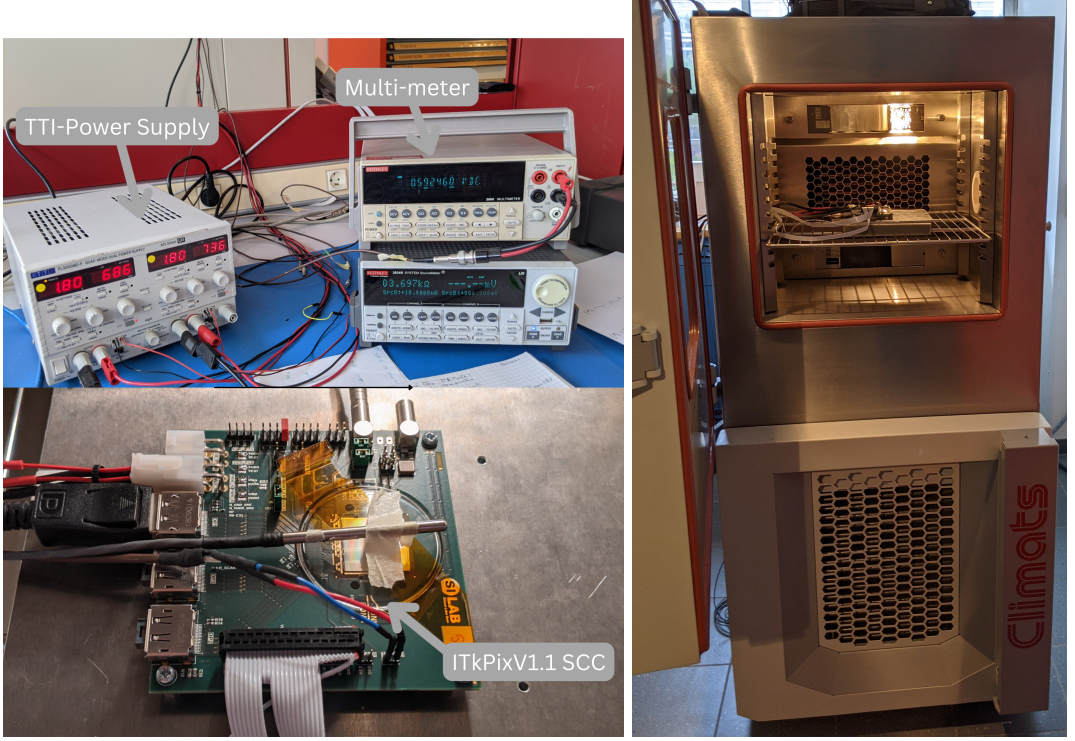}
    \caption{Experimental setup with SCC inside a climate chamber.}
\label{fig:testingsetupcompletewithnolayout}
\end{figure}

\subsection{Temperature Tests}

The ITkPix V1.1 SCC is biased at a supply voltage of 1.22~V and accommodated inside the climate chamber as in figure~\ref{fig:testingsetupcompletewithnolayout}, to allow the temperature dependency test on oscillation frequency. ROs were read out through the Keithley data acquisition system, with supply voltage $V_{dd}$  sensing lines fed back to the power supply, to eliminate any unwanted changes in bias. ROs were read with a step of $10^{\circ}\mathrm{C}$ ranging from $-40^{\circ}\mathrm{C}$ to $60^{\circ}\mathrm{C}$.

The reduction in frequency due to the temperature increase is seen in the simulation and validated by the experimental data as shown in figure~\ref{fig:roscfreqmerged}, simulation results are shown as dashed lines with experimental points for several ROs. The frequency  tends to decrease at higher temperatures, seen as a trend for all ROs also confirmed in another study in \cite{Omar_ist}. Additionally, the observed offset between simulation and experimental data arises from two main factors. Firstly, a systematic voltage mismatch is present, as the $V_{dd}$ during experimental measurements was approximately 2~mV higher than in simulations. All simulations were performed at the nominal operating voltage of 1.2~V, which serves as the reference point for evaluating voltage-induced frequency
shifts and compensation strategies. This leads to a consistent frequency offset observed across all RO variants, including Str-0 and Str-4. Secondly, there is a strength-dependent deviation, more evident in Str-0 ROs due to their smaller transistor channel widths, which are more susceptible to threshold voltage shifts. This highlights a structural dependence in the frequency response that must be considered when comparing different drive strengths to simulation results.

\subsection{Supply Voltage Tests}
\label{subsection:splvolt}
\begin{figure}[ht]
    \centering
    \includegraphics[width=1\linewidth]{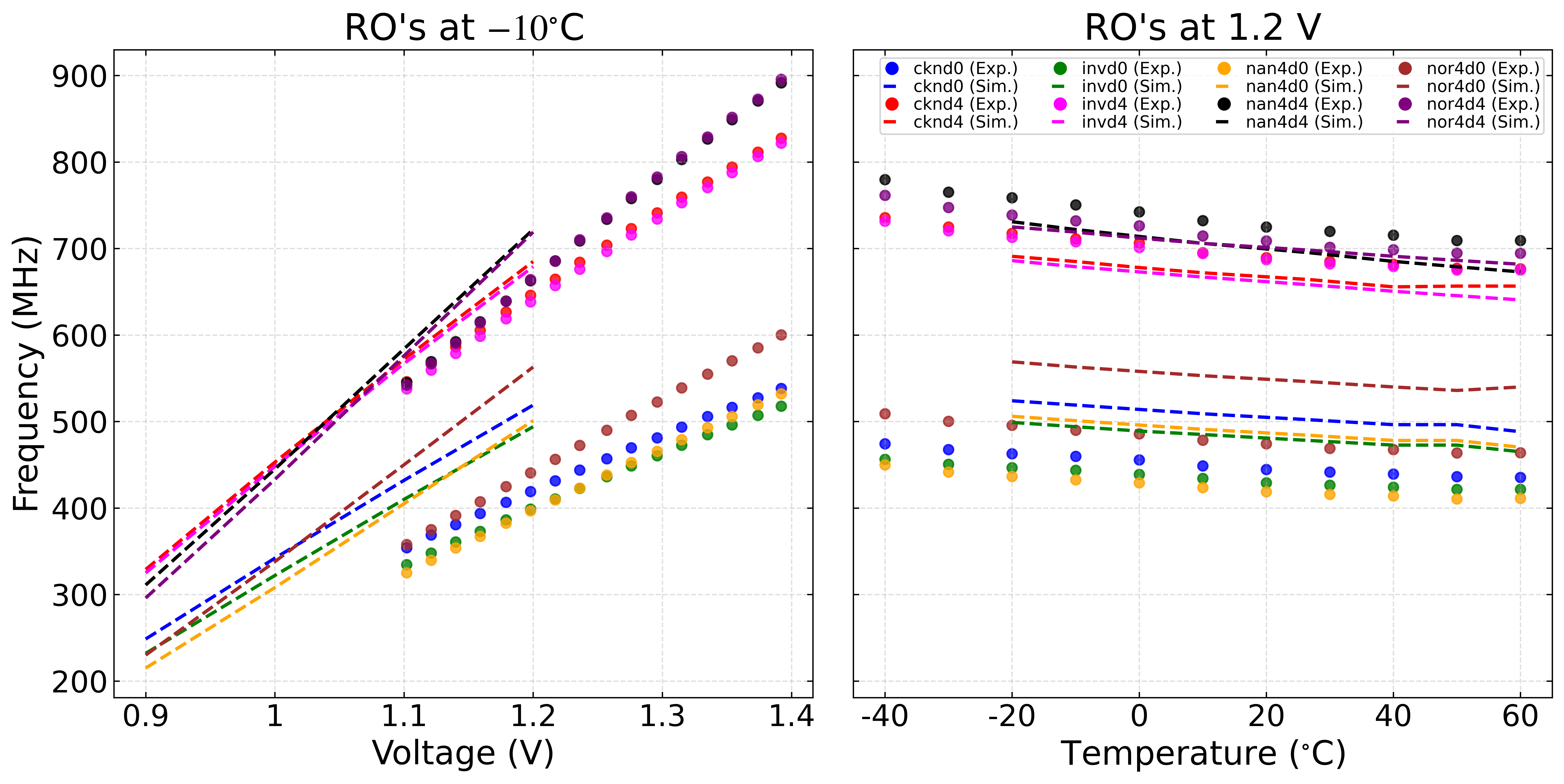}
    \caption{Comparison of RO frequency oscillation as a function
    of supply voltage and temperature. Experimental measurements (markers) are compared
    with simulation results (dashed lines). The voltage dependence is shown at $-10^{\circ}\mathrm{C}$, while the temperature dependence corresponds to simulations at
    $V_{dd}=1.2$~V.}
    \label{fig:roscfreqmerged}
\end{figure}

As the RO is embedded in the ITkPixV1.1 for radiation monitoring, supply voltage $V_{dd}$ contributes significantly to the output, as shown in equation~\ref{eq:big_eq}. Hence, frequency-to-bias dependency is important to address the frequency-to-TID accumulated over the lifetime operation as in the HL-LHC. Experimental data taken from different ROs and compared with Cadence simulations at the same temperature are shown in figure~\ref{fig:roscfreqmerged} and complemented in \cite{Omar_ist}.


\subsection{Irradiation Setup and Results}
\begin{figure}[ht]
    \centering

    \begin{subfigure}[b]{0.43\linewidth}
        \centering
        \includegraphics[width=\linewidth]{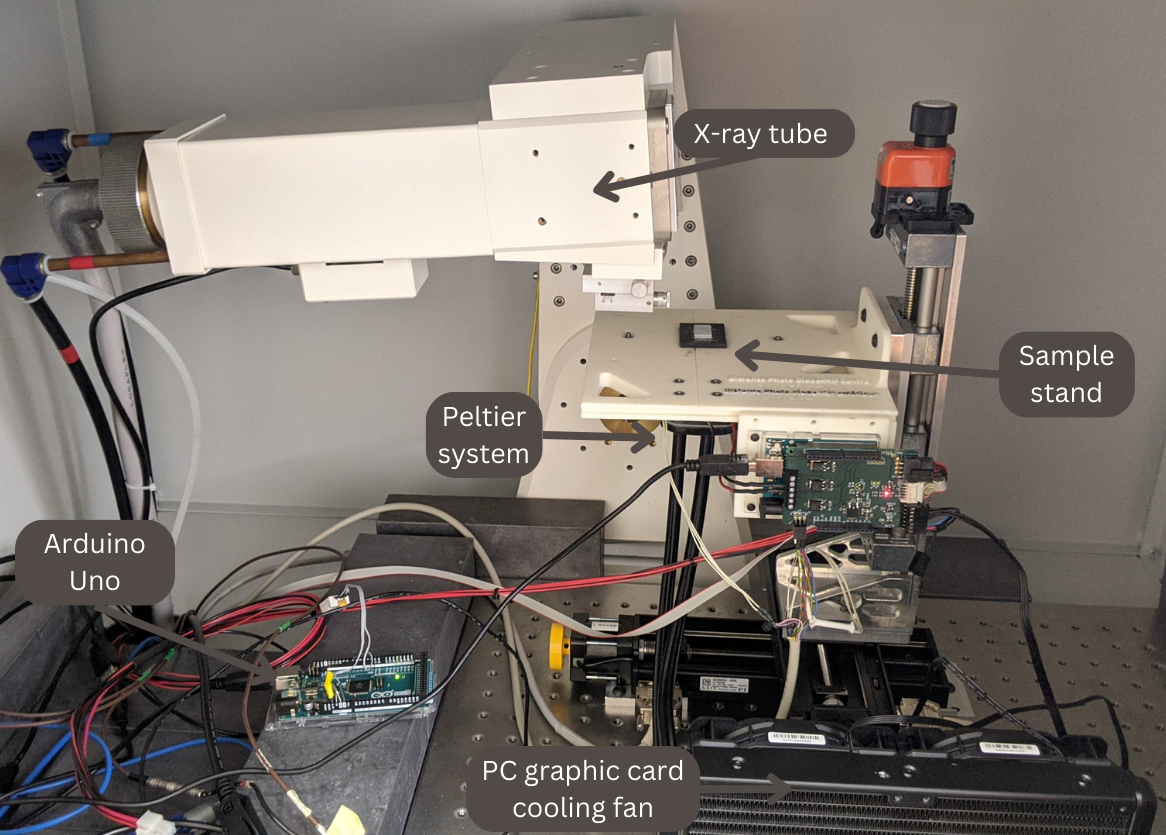}
        \caption{} 
        \label{fig:laysetupxrat}
    \end{subfigure}
    \hfill
    \begin{subfigure}[b]{0.56\linewidth}
        \centering
        \includegraphics[width=\linewidth]{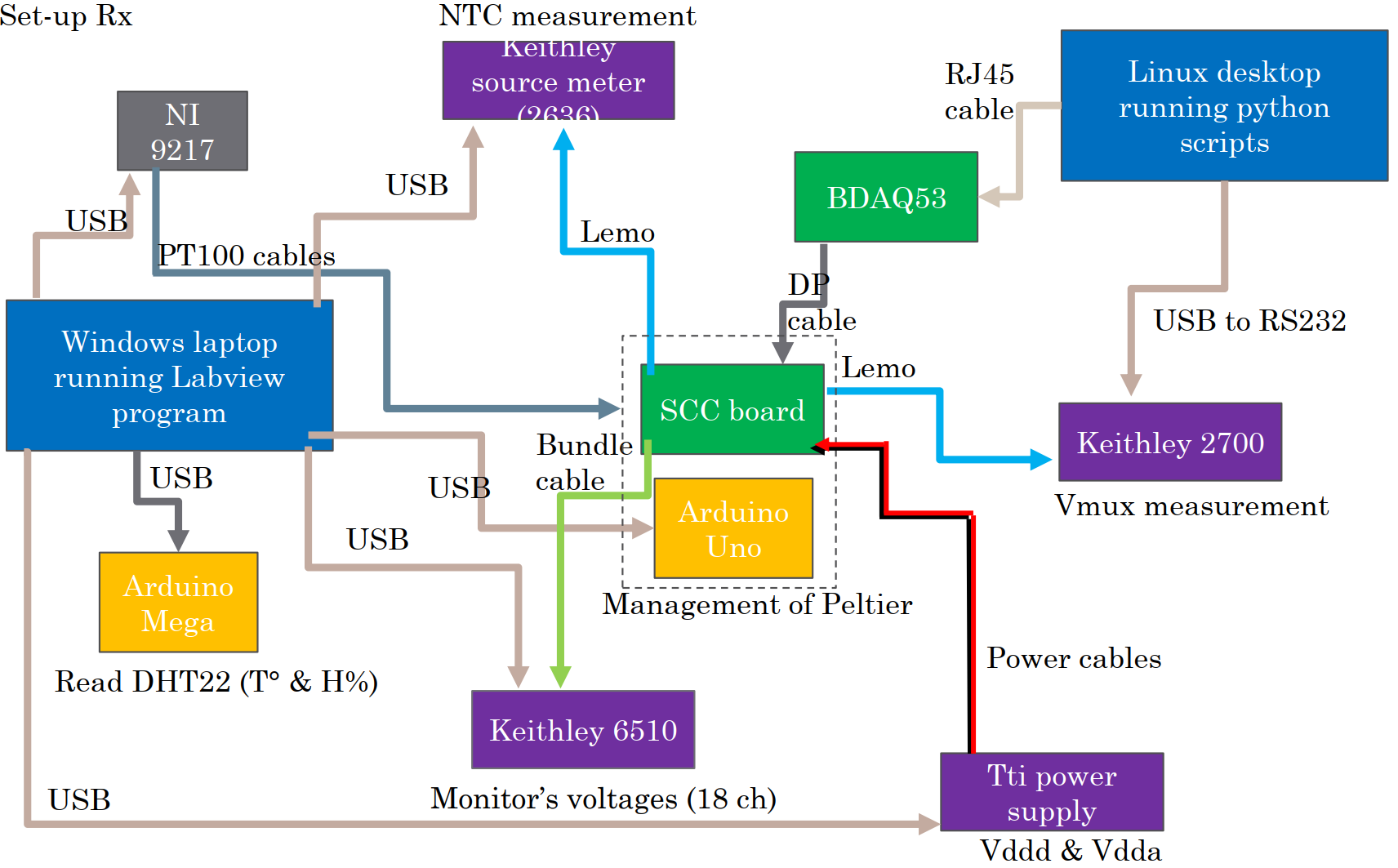}
        \caption{} 
        \label{fig:laysetupxrattt}
    \end{subfigure}

    \caption{%
    Irradiation experimental setup and measurement layout:
    (a) irradiation setup with X-ray tube generator and cooling system at IM2NP.
    (b) measuring system layout showing connections, sensors, and data acquisition systems.
    }
    \label{fig:joined_xray_setup}
\end{figure}

In figure~\ref{fig:laysetupxrat}, a complete view of the experiment is shown with the X-ray tube calibrated to deliver a 16~krad/min high dose rate (HDR) on the region of interest (ROI) for TID studies using a 2~cm beam-diameter. The measuring system used in this experiment allowed for accurate operating conditions with voltage $V_{dd}$ precisely controlled with feedback sensing lines to avoid unwanted discrepancies. Moreover, the Peltier cooling system developed at CPPM suited the experiment to provide stable temperature control against temperature changes, a complete layout of the experimental setup is seen in figure~\ref{fig:laysetupxrat} and figure~\ref{fig:laysetupxrattt}.

\begin{figure}
    \centering
    \includegraphics[width=1\linewidth]{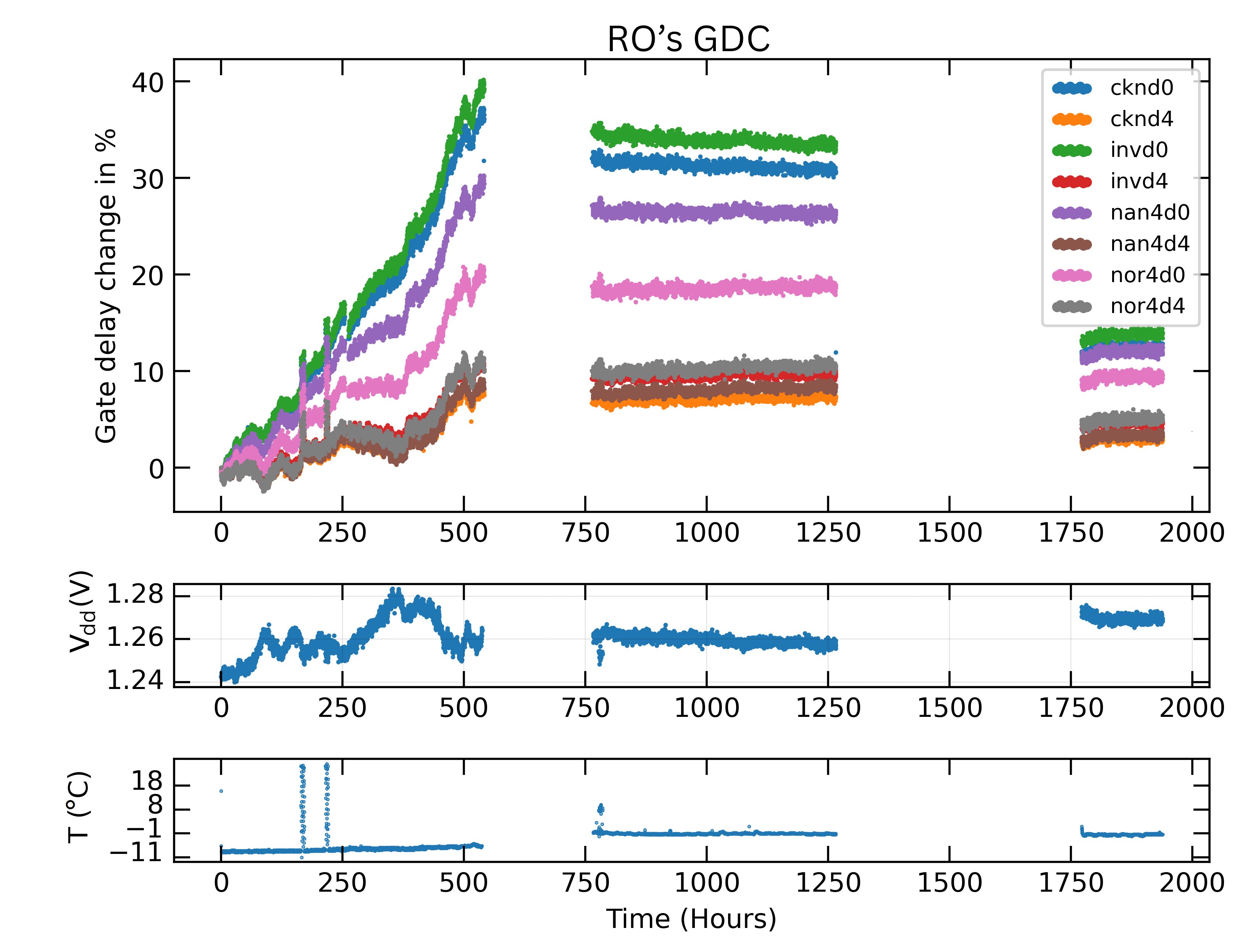}
    \caption{RO GDC evolution throughout the irradiation and annealing processes with no $V_{dd}$ corrections.}
\label{fig:firstoutput}
\end{figure}

In addition, the on-chip analog multiplexer allows selective readout of internal voltages and currents, enabling monitoring of specific signals such as RO outputs during performance characterization as illustrated in figure~\ref{fig:laysetupxrattt}. Inspection of the experiment in figure~\ref{fig:firstoutput} shows the experiment duration of 84~days, divided into three distinct periods. Irradiation occurred from the start until 520~hours, followed by the first annealing period between 750 and 1250~hours. Subsequently, ITkPixV1.1 was kept at the same temperature in a powered-off state with no data collection. Finally, a second annealing period took place from 1750 to 2000~hours.
Over the 25-day irradiation period, a total accumulated TID dose of 520~Mrad was delivered to ITkPixV1.1, with a constant bias feed and cooling temperature of $-10^{\circ}\mathrm{C}$.

\vspace{4mm}
Furthermore, the post-irradiation studies included two annealing periods with chip bias and unbias in between with temperature set at $0^{\circ}\mathrm{C}$. The first period of annealing was conducted for 20~days, and the second for 8~days to examine the effect of long-term chip annealing. As seen in figure~\ref{fig:firstoutput}, RO $T_{pd}$ expressed as gate delay change (GDC) increases rapidly through irradiation. The TID accumulated at the end of the campaign shows a maximum increase of $40\%$ in delay change increase compared to the pre-irradiation state once the experiment started at $T_{pd}^{0}$, defined as
\begin{equation}
GDC =
\frac{T_{pd} - T_{pd}^{0}}{T_{pd}^{0}}.
\end{equation}

\begin{figure}
    \centering
    \includegraphics[width=1\linewidth]{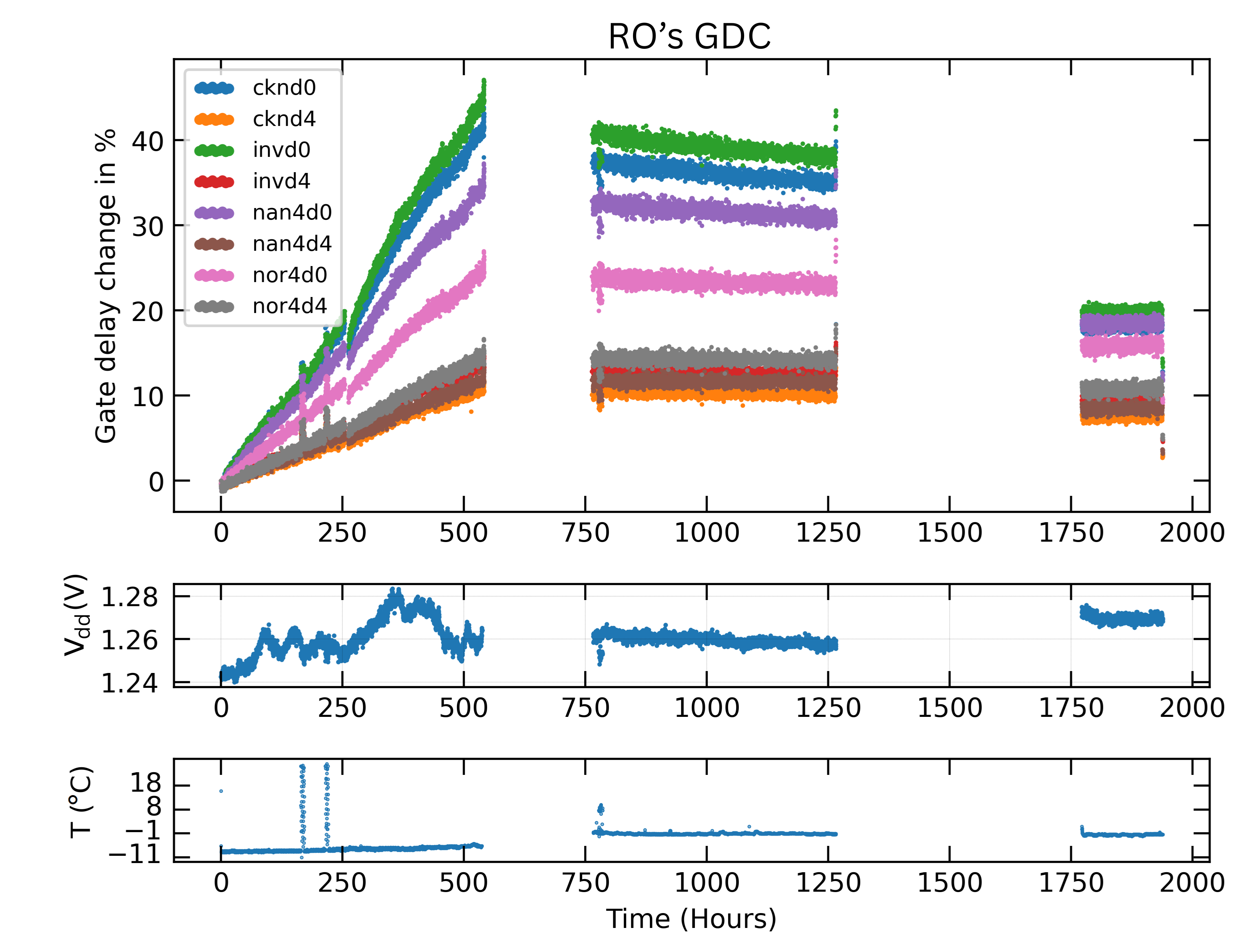}
    \caption{RO GDC evolution throughout the irradiation and annealing processes with $V_{dd}$ corrections. }
    \label{fig:correfinal}
\end{figure}

Moreover, another growth was measured in the supplied voltage $V_{dd}$ power domain. Relative to the initial conditions, $V_{dd}$ increased by 40.2~mV during irradiation. This increase reduces the propagation delay $T_{pd}$. Hence, to compensate for this unintended increase in $V_{dd}$, a voltage-dependent correction is applied to reference the data to the initial bias conditions. The correction is extracted from the pre-irradiation supply-voltage tests described in section~\ref{subsection:splvolt} and is used to adjust the measured response to the starting point of the experiment, as given by:
\begin{equation}
GDC_{\mathrm{after}} =
GDC_{\mathrm{before}} +
\frac{1}{T_{pd}^{0}}
\left(\frac{\partial T_{pd}}{\partial V_{dd}}\right)
\left(V_{dd}^{\mathrm{ref}} - V_{dd}\right),
\end{equation}
where $\partial T_{pd}/\partial V_{dd}$ represents the sensitivity of the gate propagation delay to variations in the supply voltage, and $V_{dd}^{\mathrm{ref}}$ corresponds to the nominal bias voltage at the beginning of the experiment. It is seen in figure~\ref{fig:correfinal} that the corrected GDC increases linearly with the TID accumulated dose. That is, compared to figure~\ref{fig:firstoutput}, we have a smoother performance after eliminating $V_{dd}$ inconsistencies. In addition, a $48\%$ increase in GDC was found at the end of the irradiation phase, highlighting the empirical observation of TID damage on ROs. Additionally, the correction was applied during the annealing period for proper comparison with the same initial states as in irradiation. Apart from this, annealing accelerates the radiation-induced detrapping process, which is seen in the first annealing phase as a slight drop in GDC, and a further significant drop in second annealing phase is observed for ROs. This significant drop during the second phase resulted from the chip powered-off mode. The lack of an electric field allows the thermal energy during annealing to effectively release trapped charges from the bulk defect sites, allowing the trapped charges to diffuse out of the bulk more freely.

\section{Conclusion}
\label{sec:Conc}

Experimental measurements on the ITkPixV1.1 ROs were performed and validated through Cadence simulations and an irradiation campaign. Pre-irradiation temperature and supply-voltage tests were carried out to characterize the frequency dependence of the ring oscillators under varying operating conditions. The temperature dependency test showed a reduction in frequency with increasing temperature, which was consistent for all RO flavors in the ITkPixV1.1 chip, and validated by simulation.

However, in the supply voltage study, a disagreement has been found between the simulation and experimental data. It was found that the simulated frequency corner by Cadence tends to match the Str-4 ROs, whereas, an offset is seen to Str-0 ROs. Nevertheless, the bias dependency test showed that the supply voltage contributes significantly to the RO output, and addressing the oscillation frequency to bias sensitivity is important for estimating the TID over long operation. Moreover, the irradiation campaign delivered a total of 520~Mrad, showing that $T_{pd}$ increases with the accumulated dose. 
The results demonstrate that the Str-0 ROs are significantly more affected by TID damage compared to their Str-4 counterparts. The performance degradation in Str-0 ROs scales with increased radiation exposure over time, highlighting their higher sensitivity to TID effects. Despite this, Str-0 ROs exhibit a notable ability to recover some oscillation frequency performance following annealing at high temperatures, indicating that they are more amenable to recovery after radiation-induced damage.

In contrast, Str-4 ROs display greater tolerance to TID damage, as evidenced by their relatively stable performance under similar irradiation conditions. The larger transistor channel width size in Str-4 ROs likely contributes to their improved radiation hardness.

During irradiation, an unexpected increase of $V_{dd}$ was detected during the irradiation period. Hence, an absolute correction has been applied to the results to eliminate the $V_{dd}$ fluctuations to observe purely the TID effect on GDC independent of other factors. Furthermore, the rapid increase observed after 25 days of irradiation could be compensated in the post-irradiation phase with suitably adjusted high-temperature annealing. This study clarifies a significant drop after long-term annealing, especially for the Str-0 ROs. 

 Overall, the results of the pre-irradiation campaign were confirmed by RO simulation results and assisted in understanding the TID effects. The temperature and bias dependency tests provided important insights into RO behavior under different conditions, which could be used to optimize its performance in future applications. The post-irradiation annealing studies highlighted the importance of understanding the radiation-induced detrapping process, which could be used to improve RO resilience to TID accumulation, especially during the long shutdowns in the HL-LHC phase.

\end{document}